\documentclass[aps,prapplied,reprint,superscriptaddress,amsmath,amssymb,floatfix]{revtex4-2}
\usepackage{booktabs}
\usepackage{array}
\usepackage{graphicx}
\usepackage{bm}
\usepackage{braket}
\usepackage{amsfonts}
\usepackage{color}
\usepackage{tikz}
\usepackage{booktabs}
\usepackage{array}

\begin{document}

\title{From Random Fringes to Deterministic Response: Statistical Foundations of Time-Reversed Young Interferometry}

\author{Jianming Wen}
\email{jwen7@binghamton.edu}
\affiliation{Department of Electrical and Computer Engineering, Binghamton University, Binghamton, New York 13902, USA}

\date{\today}

\begin{abstract}
Young interference is usually read as the gradual statistical accumulation of random detection events. Here we show that a time-reversed Young (TRY) geometry has a different statistical character: the fringe is not a marginal distribution of detector positions, but a conditional response indexed by a programmed source coordinate. With a fixed detector and a scanned source basis, the observable is an operational hybrid correlator between detector signal and source label. The resulting interference is deterministic at the response-function level, while noise enters only through estimation precision. We formulate this distinction using Fisher information, estimator variance, and noise scaling, clarifying why TRY naturally supports calibration, lock-in readout, null-fringe sensing, and source-plane superresolution.
\end{abstract}

\maketitle

\textit{Introduction.---}
Young's double-slit experiment is one of the clearest demonstrations of optical coherence, but the way its interference pattern is measured is intrinsically statistical~\cite{Young1804,BornWolf,GoodmanFourier,GoodmanStat,MandelWolf,Glauber1963,Glauber1963b,ScullyZubairy,SalehTeich,Loudon,Mandel1999}. The field amplitudes from different paths interfere coherently, whereas each detection event is localized and random. The familiar fringe is therefore not obtained from a single event, but from an ensemble of counts whose spatial histogram approaches the intensity distribution set by first-order coherence. In this conventional geometry, the detector coordinate is the random measurement outcome.

\begin{figure}[hbtp]
\centering
\includegraphics[width=0.75\columnwidth]{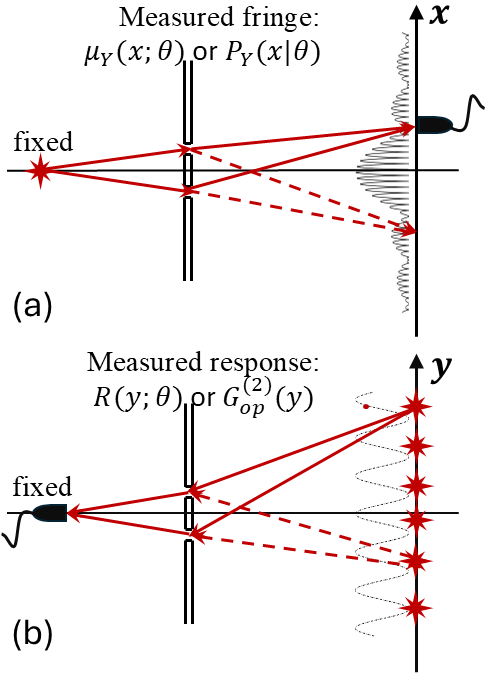}
\caption{
Statistical distinction between conventional Young interference and TRY interference. In the standard geometry, the detector coordinate $x$ is the random variable and the fringe is reconstructed from a detection histogram. In TRY, the detector is fixed and the source coordinate $y$ is a controlled label. The measured fringe is a conditional response function $R(y;\theta)$ or operational hybrid correlator $G_{\rm op}^{(2)}(y)$.
}
\label{fig:concept}
\end{figure}

Time-reversed Young (TRY) interferometry~\cite{Wen2025,WenHybrid,Wen2026a,Wen2026b,Wen2026c} changes this operational structure. The detector is fixed, while the source coordinate is scanned, switched, or otherwise programmed. The measured pattern is no longer a histogram over random detector positions. Instead, it is a source-labeled response of a fixed detector. This is not merely a reversal of arrows in an optical diagram. It changes the statistical object being measured: conventional Young interferometry measures a marginal detection distribution, whereas TRY measures a conditional optical response indexed by a controlled source coordinate, as illustrated schematically in Fig.~\ref{fig:concept}.

Figure~\ref{fig:concept}(a) shows the conventional Young configuration, where the detector-plane coordinate $x$ is the random measurement outcome and the fringe is reconstructed from spatially distributed detection events. Figure~\ref{fig:concept}(b) shows the TRY configuration, where the detector is fixed and the source coordinate $y$ is controlled; the measured fringe is therefore a conditional response rather than a marginal detection distribution.

This distinction is important because it separates two ideas that are often conflated: the coherent formation of an interference response and the statistical sampling of that response. In the standard Young experiment~\cite{Young1804,BornWolf,GoodmanFourier,GoodmanStat,MandelWolf,Glauber1963,Glauber1963b,ScullyZubairy,SalehTeich,Loudon,Mandel1999}, the two are intertwined. The optical coherence determines the ideal fringe, but the fringe is observed through stochastic spatial sampling. In TRY~\cite{Wen2025,WenHybrid,Wen2026a,Wen2026b,Wen2026c}, the coherent reciprocal response exists for each chosen source label, and noise enters through the precision with which that response is estimated. Thus deterministic fringe formation in TRY does not mean the absence of photon statistics; it means that the coordinate defining the fringe is controlled rather than randomly sampled.

Here we formulate this statistical difference in a compact estimation-theoretic language~\cite{Kay,VanTrees,Helstrom,Holevo}. We compare the two geometries at the level of optical coherence, photon-counting statistics, Fisher information~\cite{Fisher}, and the Cram\'er--Rao bound~\cite{Cramer,Rao}. For clarity, we assume paraxial, linear, reciprocal propagation; stationary illumination during each sampling interval; and independent noise between source-coordinate bins. These assumptions are sufficient to expose the central point: TRY relocates randomness from spatial pattern formation to response estimation. This relocation provides the statistical basis for source-plane coding, dwell-time optimization, null-fringe sensing, lock-in readout, and fixed-detector metrology.

\textit{Conventional Young interference as marginal detection statistics.---}
In the standard Young geometry, mutually coherent paths illuminate an extended detection plane. Let $\theta$ denote the parameter to be inferred, such as a relative phase, slit separation, source displacement, wavelength, or propagation distance. For an integration time $T$, the mean count density at detector coordinate $x$ is
\begin{equation}
\mu_Y(x;\theta)=\eta T\left\langle\hat{E}^{(-)}(x;\theta)\hat{E}^{(+)}(x;\theta)\right\rangle,
\label{eq:muY}
\end{equation}
where $\hat{E}^{(+)}$ and $\hat{E}^{(-)}$ are the positive- and negative-frequency field operators and $\eta$ is the detection efficiency. The normalized detection-position probability density is
\begin{equation}
P_Y(x|\theta)=\frac{\mu_Y(x;\theta)}{\Lambda_Y(\theta)},\quad\Lambda_Y(\theta)=
\int\mu_Y(x;\theta)\,dx.\label{eq:PY}
\end{equation}
Here $\Lambda_Y(\theta)$ is the total expected count over the detection plane. Thus $P_Y(x|\theta)$ describes the detector-position distribution conditioned on a detection event, while $\mu_Y(x;\theta)$ retains the absolute count scale.

For two interfering paths,
\begin{equation}
\hat{E}^{(+)}(x;\theta)=\hat{E}^{(+)}_1(x;\theta)+
\hat{E}^{(+)}_2(x;\theta),\label{eq:Eyoung}
\end{equation}
which gives
\begin{equation}
\mu_Y(x;\theta)=\mu_1(x;\theta)+\mu_2(x;\theta)+
2{\rm Re}\left[G^{(1)}_{12}(x;\theta)\right],
\label{eq:cohY}
\end{equation}
where
\begin{equation}
\mu_i(x;\theta)=\eta T\left\langle\hat{E}^{(-)}_i(x;\theta)
\hat{E}^{(+)}_i(x;\theta)\right\rangle,\quad
i=1,2,\label{eq:mu12}
\end{equation}
and
\begin{equation}
G^{(1)}_{12}(x;\theta)=\eta T\left\langle
\hat{E}^{(-)}_1(x;\theta)\hat{E}^{(+)}_2(x;\theta)
\right\rangle.\label{eq:G1Y}
\end{equation}
The interference term is therefore fixed by first-order coherence, while the observed fringe is obtained by sampling this intensity distribution through photon counts.

For a finite detector bin $\Delta x_j$ centered at $x_j$, define the mean count
\begin{equation}
\lambda_j(\theta)=\int_{\Delta x_j}\mu_Y(x;\theta)\,dx\simeq\mu_Y(x_j;\theta)\Delta x_j.\label{eq:lambdaj}
\end{equation}
In the shot-noise-limited regime, independent detector bins obey
\begin{equation}
n_j\sim{\rm Pois}\!\left[\lambda_j(\theta)\right],
\label{eq:poissonY}
\end{equation}
so that
\begin{equation}
\langle n_j\rangle=\lambda_j(\theta),\quad{\rm Var}(n_j)=\lambda_j(\theta).\label{eq:poissonmomentsY}
\end{equation}
The same optical intensity that defines the ideal fringe therefore also sets the local counting noise.

The information about $\theta$ is contained in how the full spatial count distribution changes with that parameter. For independent Poisson bins, the classical Fisher information~\cite{PerezDelgado2012,Matsumura2018,Luis2010,Luis2012} is
\begin{equation}
\mathcal{F}_Y(\theta)=\sum_j\frac{1}{\lambda_j(\theta)}\left[\frac{\partial \lambda_j(\theta)}{\partial\theta}\right]^2.
\label{eq:FIY}
\end{equation}
In the fine-bin limit, this becomes
\begin{equation}
\mathcal{F}_Y(\theta)=\int\frac{1}{\mu_Y(x;\theta)}
\left[\frac{\partial \mu_Y(x;\theta)}{\partial\theta}
\right]^2dx.\label{eq:FIYcontinuous}
\end{equation}
Any unbiased estimator satisfies the Cram\'er--Rao bound
\begin{equation}
{\rm Var}(\hat{\theta})\ge\frac{1}{\mathcal{F}_Y(\theta)}.\label{eq:CRBY}
\end{equation}
Equations~(\ref{eq:FIY}) and (\ref{eq:CRBY}) summarize the statistical structure of the conventional Young setup: coherence creates the intensity modulation, photon counting samples it as a random spatial distribution, and parameter sensitivity is determined by the squared fringe slope weighted by the inverse local count noise.

\textit{TRY interference as a conditional response.---}
In TRY interferometry~\cite{Wen2025,WenHybrid,Wen2026a,Wen2026b,Wen2026c}, the detector coordinate is fixed and the source coordinate is the controlled variable. Let $y(t)$ denote the source label selected at time $t$, where $y$ may represent a scanned source position, a switched source element, or a programmed source-basis index. The detector output is then naturally described by
\begin{equation}
I_D(t)=R[y(t);\theta]+\xi(t),\label{eq:modelTRY}
\end{equation}
where $R(y;\theta)$ is the mean optical response at the fixed detector for source label $y$, and $\xi(t)$ collects all fluctuations around that mean, including detector noise, source-intensity fluctuation, shot noise, timing jitter, and any residual technical drift during the sampling interval. The key point is that the coordinate defining the fringe is no longer random. It is imposed by the source program and known for every recorded datum.

Under stationary conditions within each source bin, the TRY fringe is therefore the conditional expectation
\begin{equation}
R(y;\theta)=\langle I_D(t)\,|\,y(t)=y\rangle .
\label{eq:conditionalTRY}
\end{equation}
Equivalently, one may write the operational source-resolved hybrid correlator~\cite{Wen2025,WenHybrid}
\begin{equation}
G_{\rm op}^{(2)}(y)=\frac{\langle I_D(t)\delta[y(t)-y]\rangle}{\langle \delta[y(t)-y]\rangle},
\label{eq:GopTRY}
\end{equation}
which makes explicit that the measured quantity is obtained by conditioning the detector record on the programmed source label. For a continuous source scan, $\delta[y(t)-y]$ acts as a source-coordinate selector; for a discrete source basis, it should be read as the corresponding Kronecker delta or bin indicator. Equation~\eqref{eq:GopTRY} is the statistical object naturally realized in TRY. In contrast to the conventional Young fringe, which is a marginal distribution over detector outcomes, the TRY fringe is a conditional response of a fixed detector to a controlled source coordinate.

The coherent origin of this response is still interference at the field-amplitude level. For a discrete source basis $\{y_m\}$, the field reaching the fixed detector may be written as a coherent sum
\begin{equation}
\mathcal{E}_D(y;\theta)=\sum_{m=1}^{M}
a_m\,c_m(y)\,h_D(y_m;\theta),\label{eq:EDTRY}
\end{equation}
where $a_m$ is the complex source amplitude associated with element $m$, $c_m(y)$ is the programmed weight defining the selected source coordinate or source basis, and $h_D(y_m;\theta)$ is the propagation amplitude from source element $y_m$ to the fixed detector. The corresponding response is
\begin{equation}
R(y;\theta)\propto\left|\mathcal{E}_D(y;\theta)
\right|^2=\left|\sum_{m=1}^{M}a_m\,c_m(y)\,h_D(y_m;\theta)\right|^2.\label{eq:TRYresponse}
\end{equation}
Equation~(\ref{eq:TRYresponse}) shows that TRY retains the same essential interference physics as ordinary Young interferometry: distinct amplitudes are added coherently before the modulus square is taken. The difference is not in whether interference exists, but in how that interference is interrogated statistically.

Expanding Eq.~(\ref{eq:TRYresponse}) makes this point more transparent:
\begin{eqnarray}
R(y;\theta)&\propto&\sum_{m}|a_m|^2|c_m(y)|^2 |h_D(y_m;\theta)|^2\nonumber\\
&+&\sum_{m\neq n}a_ma_n^*c_m(y)c_n^*(y)
h_D(y_m;\theta)h_D^*(y_n;\theta).\label{eq:TRYexpand}
\end{eqnarray}
The first sum gives the incoherent background contributions from individual source components, while the second sum contains the pairwise interference terms. These cross terms are weighted by the programmed source-basis coefficients $c_m(y)c_n^*(y)$, so the source label directly controls how different propagation paths interfere at the fixed detector. In this sense, TRY measures a reciprocal transfer function in a programmable source basis.

This interpretation also clarifies the role of optical coherence. In the conventional Young geometry, first-order coherence governs the visibility of a spatial fringe in the detector plane. In TRY, the same coherence governs the visibility of a source-indexed response at a fixed detector. The coherence physics is therefore not removed by time reversal; it is re-expressed through conditioning on a controlled source label. Put differently, standard Young interference resolves coherence by varying where one detects, whereas TRY resolves coherence by varying where and how one launches.

To estimate the response experimentally, the detector record is grouped according to the source label. For the bin centered at $y_k$,
\begin{equation}
\hat{R}(y_k)=\frac{1}{N_k}\sum_{t_i\in k} I_D(t_i),
\label{eq:Rest}
\end{equation}
where $N_k$ is the number of samples acquired while the source program occupies that bin. If the fluctuations are unbiased and independent between samples within the same bin, then
\begin{equation}
\langle\hat{R}(y_k)\rangle=R(y_k;\theta),\quad
{\rm Var}[\hat{R}(y_k)]=\frac{\sigma_k^2}{N_k},
\label{eq:Rvar}
\end{equation}
where $\sigma_k^2={\rm Var}[\xi(t)\,|\,y(t)=y_k]$ is the conditional noise variance in that bin. This scaling is important. It shows that averaging in TRY does not create the fringe; it only refines the estimate of a response that is already defined for each source label.

This is the precise sense in which TRY produces deterministic interference~\cite{Wen2025,WenHybrid}. The adjective ``deterministic'' refers to the response function $R(y;\theta)$, not to the absence of fluctuations in individual measurements. A single detector click or analog readout is still noisy. What changes is the role of that randomness. In the standard Young experiment, the randomness determines where the event appears in the measured fringe plane. In TRY, the source coordinate is already known, so the randomness affects only how accurately one estimates the response assigned to that coordinate.

The same logic applies in both classical and quantum operating regimes. With classical coherent or partially coherent light, $R(y;\theta)$ is the source-conditioned mean detector intensity. With single photons, $R(y;\theta)$ is the source-conditioned detection probability or count rate at the fixed detector. In either case, the operational observable is conditional rather than marginal. This is why TRY is naturally compatible with source coding, synchronous detection, and fixed-detector metrology~\cite{Wen2026a,Wen2026b,Wen2026c}: the experiment measures how a controlled source basis drives a reciprocal coherent response, rather than how random detection events populate a spatial output plane.

\textit{Fisher information and Cram\'er--Rao bound in TRY.---}
The conditional-response formulation makes the estimation problem in TRY especially transparent. Since the source label is controlled, one does not infer $\theta$ from a random spatial detection coordinate, but from how the fixed-detector response changes with the programmed source coordinate. The relevant statistical question is therefore not how accurately a fringe distribution is sampled in space, but how accurately the response function $R(y;\theta)$ can be estimated in noise.

We first assume that, after binning the data by source label $y_k$, the estimator $\hat{R}(y_k)$ is unbiased, that different source bins are statistically independent, and that the conditional noise variance is either independent of $\theta$ or locally constant over the estimation range. Under additive Gaussian noise, the covariance of $\hat{R}(y_k)$ is then $\sigma_k^2/N_k$, and the likelihood of the measured set $\{\hat{R}(y_k)\}$ is
\begin{equation}
p(\{\hat{R}(y_k)\}|\theta)\propto\prod_k\exp\!\left[
-\frac{\left(\hat{R}(y_k)-R(y_k;\theta)\right)^2}{
2\sigma_k^2/N_k}\right].\label{eq:likelihoodTRY}
\end{equation}
Under these assumptions, the classical Fisher information is
\begin{equation}
\mathcal{F}_{\rm TRY}(\theta)=\sum_k\frac{N_k}{\sigma_k^2}\left[\frac{\partial R(y_k;\theta)}{\partial\theta}\right]^2.\label{eq:FITRY}
\end{equation}
If the noise variance itself depends appreciably on $\theta$, additional Fisher-information terms involving $\partial_\theta\sigma_k^2$ must be included. The form in Eq.~(\ref{eq:FITRY}) is the local-response limit relevant to fixed-noise readout and small-parameter sensing. It has a direct physical meaning. Each source bin contributes according to three factors: the number of samples assigned to that bin, the inverse conditional noise variance, and the squared local sensitivity of the response to the parameter of interest. Unlike the standard Young geometry, where the sampling coordinate is a detector outcome, TRY allows the experimenter to decide where information is gathered and how long each source label is probed.

Any unbiased estimator $\hat{\theta}$ constructed from the binned TRY data must then satisfy the Cram\'er--Rao bound
\begin{equation}
{\rm Var}(\hat{\theta})\ge\frac{1}{\mathcal{F}_{\rm TRY}(\theta)}=\left[\sum_k\frac{N_k}{\sigma_k^2}
\left(\partial_\theta R(y_k;\theta)\right)^2
\right]^{-1}.\label{eq:CRBTRY}
\end{equation}
Equation~(\ref{eq:CRBTRY}) makes the main statistical advantage of TRY explicit: the estimation precision can be improved not only by increasing the total signal, but also by engineering the source basis so that the response slope is concentrated in low-noise regions and by allocating more dwell time to the most informative source coordinates.

It is useful to compare Eq.~(\ref{eq:FITRY}) with the Fisher information~\eqref{eq:FIY} in the conventional Young setup. In standard Young interferometry, the factor $1/\mu_Y(x_j;\theta)$ weights the squared spatial fringe slope because the signal itself determines the local counting noise. In TRY, the weighting is instead $N_k/\sigma_k^2$, which depends on the conditional noise of the fixed-detector readout and on how the source program distributes measurement time across the source bins. This is the estimation-theoretic signature of the shift from a marginal spatial distribution to a conditional response function.

The same structure persists in the Poisson-limited regime. If the fixed detector records photon counts while the source program dwells at $y_k$ for a time $T_k$, and if the mean count response is $R(y_k;\theta)$, then the counts in that bin obey
\begin{equation}
n_k\sim{\rm Pois}\!\left[T_kR(y_k;\theta)\right].
\label{eq:poissonTRY}
\end{equation}
The Fisher information becomes
\begin{equation}
\mathcal{F}_{\rm TRY}^{\rm Pois}(\theta)=\sum_k
\frac{T_k}{R(y_k;\theta)}\left[\frac{\partial R(y_k;\theta)}{\partial\theta}\right]^2.
\label{eq:FITRYPois}
\end{equation}
Equation~(\ref{eq:FITRYPois}) has the same formal structure as the Poisson result for standard Young interferometry, but its interpretation is different. The measured coordinate is no longer random; the source coordinate is imposed by design. Consequently, the free resource is not the choice of detector position after the fact, but the source basis and the dwell-time distribution chosen before the measurement.

\begin{figure}[htbp]
\centering
\begin{tikzpicture}[scale=0.95, every node/.style={font=\small}]
\draw[->,thick] (-0.2,0) -- (6.3,0) node[right] {$y$};
\draw[->,thick] (0,-0.2) -- (0,3.0) node[above] {$R(y)$};
\draw[domain=0:6,smooth,variable=\x,thick,samples=100]
plot ({\x},{1.4+1.0*cos(180*\x)});
\draw[dashed] (2.0,0) -- (2.0,2.4);
\draw[dashed] (3.0,0) -- (3.0,0.4);
\draw[dashed] (4.0,0) -- (4.0,2.4);
\node[below] at (2.0,0) {$y_1$};
\node[below] at (3.0,0) {$y_0$};
\node[below] at (4.0,0) {$y_2$};
\node[right] at (3.1,0.45) {null point};
\draw[<->] (3.0,0.4) -- (3.45,0.9);
\node[right] at (3.45,0.9) {$\partial_\theta R$};
\end{tikzpicture}
\caption{
TRY as a response-measurement problem. The fringe $R(y)$ is a deterministic response indexed by the programmed source coordinate. Noise affects the uncertainty of sampled response values, while parameter sensitivity is governed by the slope $\partial_\theta R$ weighted by the inverse noise variance. Near a biased null or low-response operating point, small perturbations can be converted into high-contrast response changes when the local parameter slope is nonzero.
}
\label{fig:response}
\end{figure}
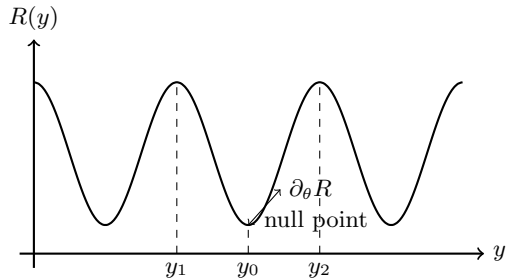

Within the Gaussian response model, the local meaning of of Eqs.~(\ref{eq:FITRY}) and \eqref{eq:FITRYPois} becomes especially clear for small parameter changes, as illustrated in Fig.~\ref{fig:response}. 
The TRY fringe is a response curve $R(y;\theta)$ indexed by the controlled source coordinate, and the useful signal is the differential change of this curve under a perturbation. 
Expanding the response around a nominal operating point gives
\begin{equation}
R(y;\theta+\delta\theta)\simeq R(y;\theta)+
\delta\theta\,\partial_\theta R(y;\theta),
\label{eq:linearTRY}
\end{equation}
so that the squared signal-to-noise ratio for detecting $\delta\theta$ is
\begin{equation}
{\rm SNR}_{\rm TRY}^2=\delta\theta^2\sum_k\frac{N_k}{\sigma_k^2}\left[\partial_\theta R(y_k;\theta)
\right]^2=\delta\theta^2\mathcal{F}_{\rm TRY}.
\label{eq:SNRTRY}
\end{equation}
Thus the Fisher information is precisely the coefficient relating the squared signal-to-noise ratio to the squared parameter shift. In TRY, the useful signal is the differential change of a deterministic response curve~\cite{Wen2026a}, while the noise is the uncertainty in reading out that curve.

A particularly favorable regime is near-null-fringe operation~\cite{Wen2026b}. Suppose the source basis is chosen so that, at a nominal parameter value $\theta_0$, the fixed detector lies near a response minimum,
\begin{equation}
R(y_0;\theta_0)\simeq R_{\rm bg},\label{eq:nullTRY}
\end{equation}
where $R_{\rm bg}$ represents a residual background, dark response, or intentionally biased operating level. If the local parameter slope is nonzero,
\begin{equation}
\partial_\theta R(y_0;\theta_0)\neq 0,
\label{eq:nonzeroslope}
\end{equation}
and the residual fluctuations are dominated by an additive noise floor $\sigma_0^2$, the local Fisher information is approximately
\begin{equation}
\mathcal{F}_{\rm null}\simeq\frac{N_0}{\sigma_0^2}
\left[\partial_\theta R(y_0;\theta_0)\right]^2.
\label{eq:FInull}
\end{equation}
Thus a biased near-null point can suppress background while retaining linear sensitivity to the parameter of interest. For an exactly symmetric intensity minimum with purely quadratic response, the linear slope in Eq.~(\ref{eq:nonzeroslope}) vanishes and the leading-order estimation problem must be treated beyond the linear model of Eq.~(\ref{eq:linearTRY}). Alternatively, Eq.~\eqref{eq:FInull} shows why TRY is naturally suited to calibration, locking, and weak-perturbation sensing. The source program can be designed so that the system operates near a highly sensitive reciprocal null, while the fixed detector monitors only the deviation from that operating point~\cite{Wen2026b}.

Taken together, Eqs.~(\ref{eq:FITRY})--(\ref{eq:FInull}) show that TRY does not eliminate statistical fluctuations; rather, it relocates them. In conventional Young interferometry, the fringe itself is built through stochastic spatial sampling. In TRY, the fringe is a deterministic response function, and statistics determine only how well that function, or a parameter encoded in it, can be estimated. This relocation is the quantitative basis for dwell-time optimization, source coding, lock-in readout, and null-fringe metrology in the TRY architecture.

The comparison developed above is further summarized in Table~\ref{tab:comparison}. The contrast is not limited to a geometrical interchange of source and detector. It extends from the level of first-order coherence to the statistical meaning of the measured fringe and the way estimation precision is quantified. In the conventional Young geometry, coherence is manifested as a spatial intensity modulation that must be sampled through random detection events across the detector plane. In TRY, the same coherent superposition is interrogated through a controlled source coordinate, so the measured quantity is a conditional response of a fixed detector. The corresponding Fisher-information and Cram\'er--Rao structures reflect this change directly: the standard Young configuration is weighted by the local statistics of spatial counts, whereas TRY is weighted by the conditional noise and sampling resources assigned to each programmed source bin.

\newcommand{\tblhead}[1]{\parbox[t]{0.18\textwidth}{\raggedright #1}}
\newcommand{\tblcolA}[1]{\parbox[t]{0.39\textwidth}{\raggedright #1}}
\newcommand{\tblcolB}[1]{\parbox[t]{0.39\textwidth}{\raggedright #1}}

\begin{table*}[t]
\caption{Comparison between conventional Young interferometry and time-reversed Young (TRY) interferometry. The two settings share the same interference physics---coherent superposition of field amplitudes---but differ in the statistical object being measured, the role of the coordinate labeling the fringe, and the way precision is controlled.}
\label{tab:comparison}
\centering
\small
\renewcommand{\arraystretch}{1.22}
\begin{tabular}{lll}
\toprule\toprule
\tblhead{Aspect} & \tblcolA{Conventional Young interferometry} & \tblcolB{Time-reversed Young interferometry (TRY)} \\
\midrule\midrule

\tblhead{Operational question}
& \tblcolA{Where is light detected in the observation plane?}
& \tblcolB{How does a fixed detector respond to a chosen source coordinate or source basis?} \\

\tblhead{Fringe coordinate}
& \tblcolA{Detector position $x$}
& \tblcolB{Controlled source coordinate or source-basis label $y$} \\

\tblhead{Statistical role of coordinate}
& \tblcolA{Random measurement outcome}
& \tblcolB{Externally programmed control parameter} \\

\tblhead{Measured quantity}
& \tblcolA{Spatial intensity or detection probability, $\mu_Y(x;\theta)$ or $P_Y(x\mid\theta)$}
& \tblcolB{Source-conditioned detector response, $R(y;\theta)$ or $G_{\rm op}^{(2)}(y)$} \\

\tblhead{Statistical type}
& \tblcolA{Marginal distribution over detector outcomes}
& \tblcolB{Conditional response indexed by source label} \\

\tblhead{Optical coherence}
& \tblcolA{First-order coherence produces a spatial interference pattern in the detector plane.}
& \tblcolB{First-order coherence produces a reciprocal interference response at a fixed detector under source-basis selection.} \\

\tblhead{Photon statistics}
& \tblcolA{Spatially resolved counts; in the shot-noise-limited regime, $n_j\sim {\rm Pois}[\mu_Y(x_j;\theta)]$.}
& \tblcolB{Source-binned fixed-detector readout; depending on the regime, additive readout noise or Poisson counts acquired during the dwell time at each source bin.} \\

\tblhead{How the fringe is obtained}
& \tblcolA{Accumulated from random detection events across space.}
& \tblcolB{Estimated by conditioning detector data on the programmed source coordinate.} \\

\tblhead{Role of noise}
& \tblcolA{Part of the spatial sampling process that reconstructs the fringe.}
& \tblcolB{Limits the precision with which an already defined response curve is estimated.} \\

\tblhead{Fisher information}
& \tblcolA{$\mathcal{F}_Y(\theta)=\sum_j \lambda_j(\theta)^{-1}[\partial_\theta \lambda_j(\theta)]^2$.}
& \tblcolB{$\mathcal{F}_{\rm TRY}(\theta)=\sum_k (N_k/\sigma_k^2)[\partial_\theta R(y_k;\theta)]^2$, or $\mathcal{F}_{\rm TRY}^{\rm Pois}(\theta)=\sum_k [T_k/R(y_k;\theta)][\partial_\theta R(y_k;\theta)]^2$.} \\

\tblhead{Cram\'er--Rao bound}
& \tblcolA{${\rm Var}(\hat{\theta})\ge \mathcal{F}_Y^{-1}$, with precision set by spatial fringe slope and local counting statistics.}
& \tblcolB{${\rm Var}(\hat{\theta})\ge \mathcal{F}_{\rm TRY}^{-1}$, with precision set by response slope, conditional noise, and dwell-time allocation.} \\

\tblhead{Optimization handle}
& \tblcolA{Detector-plane sampling, aperture, and spatial integration strategy.}
& \tblcolB{Source coding, source-basis design, dwell-time distribution, and operation near a controlled null.} \\

\tblhead{Natural applications}
& \tblcolA{Spatially resolved interference measurement and imaging.}
& \tblcolB{Fixed-detector metrology, source-plane coding, null-fringe sensing, lock-in readout, and programmable reciprocal-response measurements.} \\

\bottomrule\bottomrule
\end{tabular}
\end{table*}
Table~\ref{tab:comparison} shows that TRY preserves the interference mechanism of the Young geometry while shifting the statistical burden of the measurement from spatial pattern formation to conditional response estimation.

\textit{Physical interpretation.---}
The statistical contrast between conventional Young interferometry and TRY can be summarized in an operationally precise way. In the standard Young geometry, the experiment asks where the optical field is detected. The detector coordinate is therefore the measurement outcome, and the observed fringe is obtained only after many random events populate the detection plane. In TRY, the experiment asks how a fixed detector responds when the source coordinate or source basis is selected. The coordinate defining the fringe is therefore not an outcome but a control parameter. What is measured is not a spatial probability distribution, but a source-conditioned response function.

This distinction should not be confused with a change in the underlying interference mechanism. In both geometries, the response originates from coherent superposition of field amplitudes. The same reciprocity and linear propagation that govern ordinary Young interference also govern TRY. What changes is the statistical role assigned to the coordinate. In the conventional geometry, coherence is read out by resolving the detector plane. In TRY, coherence is read out by interrogating the reciprocal optical response through a programmed source coordinate. One geometry scans where the light is received; the other scans where and how the light is launched.

This viewpoint clarifies what is meant by deterministic interference in TRY. The word ``deterministic'' refers to the existence of a well-defined response curve $R(y;\theta)$ for every chosen source label $y$. It does not mean that individual detector outputs are noise free. A single count or analog sample can still fluctuate because of shot noise, detector noise, technical drift, or source instability. However, these fluctuations no longer determine the coordinate at which the event contributes to the fringe. That coordinate is already specified by the source program. Randomness therefore enters only through the uncertainty of the measured response value associated with a known source label.

This relocation of randomness has a direct physical consequence. In conventional Young interferometry, pattern formation and noise are intertwined: the fringe is reconstructed from stochastic spatial sampling, so the statistics are part of the pattern-building process itself. In TRY, the fringe is already encoded in the reciprocal transfer function of the system. Measurement serves to estimate that response rather than to discover where events landed in space. The experiment is therefore closer in spirit to probing a system response function than to recording a passive output distribution.

The same interpretation applies in both classical and quantum regimes. With classical coherent or partially coherent light, TRY measures a source-conditioned mean intensity at a fixed detector. With single photons, it measures a source-conditioned detection probability or count rate. In either case, the operational observable is conditional rather than marginal. This is why TRY naturally accommodates source coding, synchronous or lock-in detection, dwell-time optimization, and null-response operation. Once the source label is treated as the controlled coordinate, the experiment acquires a built-in handle for shaping where information is concentrated and how it is extracted.

From this perspective, the practical appeal of TRY becomes clear. A fixed detector can monitor a programmable reciprocal interference response without requiring a spatially resolved detector array. The source basis can be engineered to emphasize steep response regions, suppress unwanted backgrounds, or place the system near a controlled null. These possibilities do not arise because TRY abandons photon statistics, but because it reorganizes them. Standard Young interferometry samples a spatial fringe produced by stochastic detection outcomes. TRY estimates a deterministic, source-indexed interference response in the presence of noise. That statistical reorganization is the conceptual bridge between the familiar Young experiment and the applied sensing and metrological capabilities of the TRY architecture.

\textit{Conclusion.---}
Time-reversed Young interferometry realizes a different measurement architecture from the conventional Young experiment. Instead of reconstructing a fringe from position-resolved detection events, it interrogates a fixed detector through a controlled source coordinate and thereby accesses a source-conditioned response. This shift leaves the interference physics intact but changes the statistical object and the way measurement uncertainty enters the problem.

Framed in terms of coherence, counting statistics, Fisher information, and the Cram\'er--Rao bound, TRY emerges as a reciprocal response-measurement scheme whose performance can be shaped through source coding, dwell-time allocation, and operation near sensitive working points. This perspective gives deterministic source-plane interference a clear statistical meaning and highlights its value for fixed-detector sensing, calibration, and precision metrology.

\begin{acknowledgments}
The author acknowledges helpful discussions with Drs. Sidong Lei and Yanhua Zhai. This work was partially supported by Binghamton University through startup funds and NSF ExpandQISE 2329027.
\end{acknowledgments}

\end{document}